\documentclass[prb,twocolumn,showpacs,amsmath,amssymb]{revtex4}

\usepackage{graphicx}
\usepackage{amssymb,amsmath}
\usepackage{dcolumn}
\usepackage{bm}
\usepackage{color}

\begin{document}

\title{Effect of the TE-TM splitting on the topological stability of Half-vortices in spinor exciton-polariton condensates}

\author{H. Flayac}
\affiliation{LASMEA, UMR CNRS-Universit\'e Blaise Pascal 6602, 24
Avenue des Landais, 63177 Aubi\`ere Cedex France}

\author{I.A. Shelykh}
\affiliation{Physics Department, University of Iceland, Dunhaga-3,
IS-107, Reykjavik, Iceland and St. Petersburg State Polytechnical
University, Polytechnicheskaya 29, 195251, St. Petersburg,
Russia}

\author{D.D. Solnyshkov}
\affiliation{LASMEA, UMR CNRS-Universit\'e Blaise Pascal 6602, 24
Avenue des Landais, 63177 Aubi\`ere Cedex France}

\author{G. Malpuech}
\affiliation{LASMEA, UMR CNRS-Universit\'e Blaise Pascal 6602, 24
Avenue des Landais, 63177 Aubi\`ere Cedex France}

\begin{abstract}
Half vortices have been recently shown to be the elementary topological defects
supported by a spinor cavity exciton-polaritons condensates with spin anisotropic interactions (Y. G. Rubo, \emph{Phys. Rev. Lett.} \textbf{99}, 106401 (2007)).
A half vortex is composed by an integer vortex for one circular component of the condensate, whereas
the other component remain static. We analyze theoretically the effect of the splitting between TE and TM polarized eigen modes on
the structure of the vortices in this system.
For TE and TM modes, the polarization states depend on the direction of propagations of particles and
imposes some well defined phase relation between the two circular component.
As a result elementary topogical defects in this system are no more half vortices but integer vortices correspond to an
integer vortex for both circular components of the condensate. The intrinsic life time of half vortices is given
and the texture of a few vortex states is analyzed.
\end{abstract}

\maketitle

\section{Introduction}
Interactions between quantum particles lie behind a number of
intriguing phenomena in the field of condensed matter physics. Being
treated within mean field approximation, for a system of interacting
bosons they result in a non-linear term in the Gross-Pitaevskii equation,
which is currently routinely used for the description of
dynamics of Bose-Einstein Condensates (BECs) of cold atoms \cite{GPBEC}. A similar equation,
known as non-linear Schr\"{o}dinger equation, is widely used in non-linear
optics for description of such phenomena as self-focusing of laser
beams and propagation of solitons \cite{Kivshar}.

The fields of BEC and non-linear optics meet each other in the context
of planar semiconductor microcavities: the mesoscopic
objects designed to enhance the light-matter interaction. A microcavity
consists of a pair of distributed Bragg mirrors confining an
electromagnetic mode and one or several Quantum Wells (QWs) with an excitonic
resonance, which are placed at the antinodes of the electric field $\overrightarrow{E}$. In strong
coupling regime, where coherent exciton-photon interaction
overcomes the damping provided by the finite lifetime of excitons and
cavity photons, a new type of elementary excitations, called exciton-polaritons (or cavity
polariton), appears in the system. The polaritons are a mixture of material excitations
(excitons) with light (photons).

The hybrid nature of polaritons gives them a set of
peculiar properties. First, at relatively small densities, polaritons
exhibit bosonic properties \cite{StimScatt}. Second, due to the
presence of a photonic component, the effective mass of the
polaritons is extremely small ($10^{-4}-10^{-5}$ of the free
electron mass), while the presence of an excitonic component makes
possible efficient polariton-polariton and polariton-phonon
interactions. These properties make possible polariton
Bose condensation \cite{Imamoglu} suggested  more than 10 years ago, up to high temperatures \cite{Malpuech02, Zamfirescu}.
The simulations have shown that relaxation of polaritons can become faster than their radiative lifetime, allowing the formation
of a quasi-equilibrium polariton gas. These predictions have been confirmed by the recent observation of polariton condensation \cite{Richard,Kasprzak,Snoke,Yamamoto,Baumberg,Christmann,Kasprzak08,Wertz} and the demonstration of the thermodynamic regime
\cite{Kasprzak08,Wertz} where the behavior of the polariton gas is well described by its thermodynamic variables
(temperature and chemical potential).
The next step after the observation of the condensation itself is to study the dynamical properties and
the specificities of polariton condensates. One of the important properties is superfluidity.
The phase transition expected for 2D polaritons is rather a Berezinskii-Kosterlitz-Thouless (BKT) transition toward a superfluid state \cite{Malpuech03}, and not the BEC. Such a phase
transition has not been immediately observed in CdTe and GaN based structures, because of the presence of a structural disorder, which has led to the formation of an Anderson Glass phase \cite{Malpuech07} or to the condensation in a single in-plane
potential trap \cite{Sanvitto09}.
Only in a cleaner GaAs-based sample some signatures of BKT phase transition have been reported \cite{Yamamoto08}. If this observation is confirmed, it would rule out the claims that no superfluid behavior can be achieved in a system of particles showing a finite
life time \cite{Marzena,Carusotto07}. Another way to excite a superfluid flow of polaritons is to properly design a resonant
excitation experiment, as described theoretically  \cite{Carusotto04,Solnyshkov07} and recently evidenced experimentally \cite{Amo09}.
In this framework the study of fundamental properties of polariton vortices is of a strong interest.
On one side, the BKT transition between normal
and superfluid states in two-dimensional system is closely connected
with the formation of topological defects (vortex-antivortex pairs).
On the other side, the recent growth of the experimental activity devoted to cavity polariton condensates
opened a race to the observation of exotic phenomena.
Observation of a vortex pinned to a defect in a disordered cavity has been reported \cite{Lagoudakis}, whereas
the formation of a lattice of vortices in a potential trap has been predicted theoretically in the scope of
a Ginzburg-Landau model \cite{Keeling}. In these two works the peculiar spin structure of polaritons was not
taken into account. In fact, only one theoretical work did study vortex states in homogeneous spinor
polariton condensates \cite{Rubo}. In this work Y. Rubo shows that elementary polariton vortex states are
the so-called "half-vortices". They are characterized by a half-quantum change of the phase of the condensate, i.e.
the phase of the wave function is changed by $\pm\pi$ after encircling the point of singularity.
This analysis however was not considering the facts that polariton eigen states in a microcavity are normally
TE or TM polarized \cite{Panzarini}, with a finite energy splitting between these two states (TE-TM splitting).

In the present work we therefore consider the impact of the TE-TM splitting on polariton vortices. We first show that using the basis of circularly polarized states (contrary to the basis of linearly polarized states used by Rubo \cite{Rubo})
allows to describe a half vortex as one vortex
for one circular component, whereas the other circular component remains immobile.
We then show that the TE-TM splitting couples the half-vortices of opposite circularity, which cease to be the stationary solutions of the spinor Gross-Pitaevskii equations.
The elementary (stationary) excitation of the condensate with the TE-TM splitting is composed by one vortex of each circular component.
This result does not mean that the half-vortices cannot be observed experimentally. It is however
a state which should decay in time and therefore should not be used for the calculation of the critical temperature of the BKT
phase transition. In the second section we present in details the spin structure of cavity polaritons.
In the third section the polarization structure of polariton vortices is analyzed. Results
and discussions are presented in the fourth section. The fifth section draws the main conclusions.

\section{Spin structure of cavity polaritons}

An important peculiarity of cavity polaritons is linked with their
spin structure. Like other bosons, polaritons exhibit an
integer spin, inherited from spins of excitons and photons. In QWs
the lowest energy level of a heavy-hole (having a spin $S_z=3/2$)
lies typically lower than any light-hole level ($S_z=1/2$) and thus
the entire exciton spin in a QW has projections $S_z=\pm2,\pm1$ on
the structure growth axis. The states with $S_z=\pm2$ are not
coupled to light and thus do not participate in polariton formation.
As they are split-off in energy, normally they can be neglected
while considering polariton dynamics \cite{LuisSSC}. On the
contrary, states with $S_z=\pm1$ form the optically active polariton
doublet and can be created by $\sigma_+$ and $\sigma_-$ circularly
polarized light, respectively. Thus, from the formal point of view,
the spin structure of cavity polaritons is analogical to spin
structure of the electrons (both being two-level systems), which
permits to introduce the concept of a pseudo-spin vector $\overrightarrow{S}$ for
the description of their polarization dynamics \cite{KirillPRL}. The
latter is determined as the coefficient of the decomposition of the
$2\times2$ spin density matrix $\rho$ of polaritons on a set
consisting of the unity matrix \textbf{I} and three Pauli matrices
$\sigma_{x,y,z}$.

\begin{equation}
\rho=\frac{N}{2}\textbf{I}+\overrightarrow{S}\cdot\overrightarrow{\sigma}
\end{equation}
$N$ is the total number of particles. The orientation of the pseudo-spin completely determines the
polarization of the emission from a microcavity. According to a
generally accepted convention, orientation of the pseudo-spin along
$z$-axis corresponds to circular polarized emission, while
pseudo-spin lying in $x-y$ plane corresponds to linear polarized
emission.

The spin dynamics of cavity polaritons has become a field of intense
research since 2002 \cite{LolaPRL}. It is governed by two factors.
First, at $\overrightarrow{k}\neq0$ there is an effective in-plane magnetic
field which results in the pseudo-spin rotation manifesting in the oscillations
of the polarization degree of photoemission in the time domain. It is
well known that due to the long-range exchange interaction between
the electron and hole, for excitons having non-zero in-plane
wave-vectors, the states with dipole moment oriented along and
perpendicular to the wave vector are slightly different in energy
\cite{Maialle}. In microcavities, the TE-TM splitting of
polariton states is greatly amplified due to the exciton coupling
with the cavity mode, which is also split in TE and TM
polarizations \cite{Panzarini}. An important feature of the
effective magnetic field generated by the TE-TM splitting is the
dependence of its direction on the direction of the wave-vector : it
is oriented in the plane of the microcavity and makes a double angle
with the $x$-axis in the reciprocal space :
\begin{equation}
\vec{\Omega}_{eff}(k)\sim\mathbf{e}_x\text{cos}(2\phi)+\mathbf{e}_y\text{sin}%
(2\phi). \label{OmegaEff}
\end{equation}
This peculiar link between the orientation of the effective
magnetic field and polariton wave-vector leads to remarkable effects
in the real-space dynamics of the polarization in quantum
microcavities, including the optical spin Hall effect \cite{OSHE},
possible formation of polarization patterns \cite{Langbeincross},
and creation of polarization vortices \cite{LiewVortex}.

Second, polariton-polariton interactions are known to be spin-
anisotropic. Since the exchange interaction plays a major role, the
interaction of polaritons with parallel spin projections on the
structure growth axis is much stronger than that of polaritons with
antiparallel spin projections \cite{Combescot}. This leads to a
mixing of linearly polarized polariton states, manifesting itself
in remarkable non-linear effects in polariton spin relaxation, such
as self-induced Larmor precession and inversion of linear
polarization upon scattering \cite{Krizhanovskii}.

In the domain of polariton BEC, spin properties of cavity polaritons
play a major role. It was argued that under unpolarized non-resonant pump the transition to phase-coherent states should be
accompanied by spontaneous appearance of a linear polarization in
the emission from the ground state. Consequently, linear
polarization can be considered as an experimentally measurable order
parameter of the polariton BEC \cite{Laussy}.

It is well known that the BKT transition between normal
and superfluid states in two-dimensional systems is closely connected
with the formation of the topological defects (vortex-antivortex pairs). It
is thus of a crucial importance to understand the structure and
polarization properties of vortices in the homogeneous polariton
condensates.

\section{Polarization vortices in spinor polariton condensates}

To the best of our knowledge, up to the present time, there exists only
one theoretical work regarding vortices in the context of the
spinor polariton condensation \cite{Rubo}. In this pioneer paper the
polarization structure of the polariton vortices was analyzed and
the existence of peculiar half-vortices was predicted
\cite{Vortexsupercond}. It was shown that contrary to the case of a
normal vortex in scalar superfluid, the particle density differs
from zero in the center of a half-vortex. Besides, these objects
have been predicted to possess a peculiar spatial dependence of the
polarization: it is circular in the center of the core and becomes
linear at large distances from it. The energy required to create a half-vortex
is twice smaller than the one required to create a normal vortex, because
only one half of the total fluid mass is rotating. As a result, the existence of half vortices
as stationary stable states divides by 2 the critical temperature of the BKT
phase transition as discussed in Ref.\onlinecite{Rubo}.

However, the effects of the in-plane effective magnetic fields of various
nature, in particular of the TE-TM splitting, on the structure of the
polarization vortices was neglected in this seminal work. As we shall see
below, these fields can have drastic effects on the structure of
polarization vortices. Besides, in our opinion, the choice of the
basis of linear polarizations used in Ref.\cite{Rubo}
hindered the clear physical understanding of the physical origin of
the half-vortices. In the present manuscript we revise and extend the
results of Y. Rubo, accounting for non zero TE-TM splitting of a
polariton doublet and using the basis of circular polarizations,
which makes the obtained results much more transparent.

The Hamiltonian of an interacting polariton system written in the
basis of circular polarized states reads :

\begin{eqnarray}
\widehat{H}=\widehat{H}_0+\widehat{H}_{int}=\int\left[\overrightarrow{\psi}^\dag\widehat{\textbf{T}}(-i\nabla)\overrightarrow{\psi}-
\mu\left(\overrightarrow{\psi}^\dag\overrightarrow{\psi}\right)\right]d\textbf{r}\\
\nonumber+\int\left[\frac{\alpha_1}{2}\left(|\psi_+|^4+|\psi_-|^4\right)+
\alpha_2|\psi_+|^2|\psi_-|^2\right]d\textbf{r}
\end{eqnarray}

where $\psi_\pm$ are the field operators for right and left circular
polarized polaritons, $\overrightarrow{\psi}=(\psi_+,\psi_-)^T$, the
coefficients $\alpha_1$ and $\alpha_2$ describe the interaction
between the polaritons with same and opposite circular polarizations
\cite{Kirill}, $\mu$ is the chemical potential determined by the
condensate density at infinity. The parameters we use are connected
with those introduced in Ref.\onlinecite{Rubo} in the following way :

\begin{eqnarray}
U_0=\alpha_1,\\
U_1=(\alpha_1-\alpha_2)/2
\end{eqnarray}

The tensor of the kinetic energy reads :
\begin{eqnarray}
\widehat{\textbf{T}}(-i\nabla)=\left(\begin{array}{cc}\widehat{H}_0(-i\nabla)&\widehat{H}_{TE-TM}(-i\nabla)\\
\widehat{H}_{TE-TM}^\dag(-i\nabla)&\widehat{H}_0(-i\nabla)\end{array}\right)
\end{eqnarray}
where the diagonal terms $\widehat{H}_0$ describe the kinetic energy of
lower cavity polaritons, and the off-diagonal terms
$\widehat{H}_{TE-TM}$ correspond to the longitudinal-transverse
splitting, mixing opposite circular polarized components. In our
further considerations we will adopt the effective mass approximation,

\begin{eqnarray}
\widehat{H}_0=-\frac{\hbar^2}{2m^*}\nabla^2;
\end{eqnarray}
\begin{eqnarray}
\widehat{H}_{TE-TM}=\beta\left(\frac{\partial} {\partial
y}+i\frac{\partial}{\partial x}\right)^2 \label{HTETM}
\end{eqnarray}
where $m^*$ is the effective mass of cavity polaritons. The Eq.
\ref{HTETM} is the simplest form of the Hamiltonian providing the
correct symmetry of the effective magnetic field given by the
expression \ref{OmegaEff} \cite{Maialle,Pikus}. The dependence of
the absolute value of this field on the wave number is taken to be
quadratic, which corresponds well to the effective mass
approximation we are using in the current paper. $\beta$ is a constant,
characterizing the strength of the TE-TM splitting which can be
expressed via the longitudinal and transverse polariton effective masses
$m_l$ and $m_t$ :

\begin{equation}
\beta=\frac{\hbar^2}{4}\left(\frac{1}{m_l}-\frac{1}{m_t}\right)
\label{beta}
\end{equation}

Within the framework of mean-field approximation at $T=0$, the
dynamics of the spinor polariton superfluid can be completely described by
a set of 2 coupled Gross-Pitaevskii equations
\cite{ShelykhPRL2006}, which in the basis of circular polarized
states reads :

\begin{widetext}
\begin{eqnarray}
i\hbar\frac{\partial}{\partial t}\left(\begin{array}{cc}\psi_+\\ \psi_-\end{array}\right)=\left(\begin{array}{cc}
   -\frac{\hbar^2}{2m^*}\nabla^2-\mu+\alpha_1|\psi_+|^2+\alpha_2|\psi_-|^2 & \beta\left(\frac{\partial}
{\partial y}+i\frac{\partial}{\partial x}\right)^2\\
 \beta\left(\frac{\partial}
{\partial y}-i\frac{\partial}{\partial x}\right)^2& -\frac{\hbar^2}{2m^*}\nabla^2-\mu+\alpha_1|\psi_-|^2+\alpha_2|\psi_+|^2 \end{array}\right)
\left(\begin{array}{cc}\psi_+\\ \psi_-\end{array}\right)
\label{GP}
\end{eqnarray}
\end{widetext}
where the chemical potential is
$\mu=\left(\alpha_1+\alpha_2\right)n_\infty/2$ with
$n_\infty=|\psi_+(\infty)|^2+|\psi_-(\infty)|^2$ being the condensate
density far away from the vortex core. Rescaling the variables
$\psi_{\pm}\rightarrow\left(\mu/(\alpha_1+\alpha_2)\right)^{1/2}\psi_{\pm}$, $\textbf{r}\rightarrow
\left(\hbar^2/(m^*\mu)\right)^{1/2}\textbf{r}$ and $t\rightarrow\left(\hbar/\mu\right)t$, one can represent the system
\ref{GP} in the following dimensionless form :

\begin{widetext}
\begin{eqnarray}
i\frac{\partial}{\partial t}\left(\begin{array}{cc}\psi_+\\ \psi_-\end{array}\right)=\left(\begin{array}{cc}
   -\frac{1}{2}\nabla^2-1+A_1|\psi_+|^2+A_2|\psi_-|^2 & \chi\left(\frac{\partial}
{\partial y}+i\frac{\partial}{\partial x}\right)^2\\
 \chi\left(\frac{\partial}
{\partial y}-i\frac{\partial}{\partial x}\right)^2& -\frac{1}{2}\nabla^2-1+A_1|\psi_-|^2+A_2|\psi_+|^2 \end{array}\right)
\left(\begin{array}{cc}\psi_+\\ \psi_-\end{array}\right)
\label{GPDimensionless}
\end{eqnarray}
\end{widetext}
where $A_{1,2} =\alpha_{1,2} /(\alpha_1+\alpha_2)$ and $\chi=\beta m^*/\hbar^2$.

Let us start our analysis from the simplest case, where the TE-TM
splitting can be neglected, $\chi=0$. This case has been considered
already in Ref.\onlinecite{Rubo}, but we feel that it will be
instructive to re-examine it using the basis of the circular
polarized states, because the final result is more transparent.

Equations \ref{GPDimensionless} allow a time-independent solution, which can be represented in the following form:

\begin{eqnarray}
\overrightarrow{\psi}=\left(\begin{array}{cc} \psi_+(r,\theta)\\
\psi_-(r,\theta)\end{array}\right)=\left(\begin{array}{cc}
f_{+}(r)e^{il_+\theta}\\ f_{-}(r)e^{il_-\theta}\end{array}\right)
\label{PsiVort}
\end{eqnarray}
where $(r,\theta)$ are the polar coordinates. Due to the conservation of
the z-component of the spin by polariton-polariton interactions, the
winding numbers of the two circular polarized components $l_{\pm}$
are independent. Rewriting Eq.\ref{PsiVort} in the basis of linear
polarized components $\psi_\pm=2^{-1/2}(\psi_X\pm i\psi_Y)$, one
easily obtains the relation between our winding numbers $l_\pm$ and
those of Ref.\onlinecite{Rubo} :

\begin{eqnarray}
k=\frac{l_+-l_-}{2},\\
m=\frac{l_++l_-}{2}.
\end{eqnarray}

The situation describing a half-vortex corresponds to the case where for one circular polarized component the winding number is zero (say $l_+=0$), while for the other one it is ($l_-=+1$). Radial wave functions $f_{\pm}$ satisfy the following set of equations:

\begin{eqnarray}
f_ + ^{''}+f_ + ^{'}+\left(2-2A_{1}f_+^2-2A_{2}f_-^2-\frac{{l_ + ^2}}{{{r^2}}}\right)f_+=0\\
f_ - ^{''}+f_ - ^{'}+\left(2-2A_{1}f_-^2-2A_{2}f_+^2-\frac{{l_ - ^2}}{{{r^2}}}\right)f_-=0
\end{eqnarray}

Which corresponds to Eqs.10 of the Ref.\onlinecite{Rubo}, if one
puts $f_\pm=2^{-1/2}(f\pm g)$.

In the simplest case, when the circular polarized components do not
interact ($A_{2}=0$), the half-vortex with $l_+=0$, $l_-=1$ corresponds
to a homogeneous distribution of $\sigma_+$ component and a simple
vortex in $\sigma_-$. Clear enough, in the center of such a half-vortex the
density is non-zero (due to the $\sigma_+$ component) and polarization
is circular, since the density of the $\sigma_-$ component is zero in the
center of the vortex. Moving from the center of the vortex changes polarization from circular to linear in a
continuous manner.

Now let us consider a more interesting case where
$\chi\neq0$. The terms associated with the TE-TM splitting
rewritten in polar coordinates read:

\begin{widetext}
\begin{eqnarray}
\left(\frac{\partial}{\partial y}\pm i\frac{\partial}{\partial
x}\right)^2 = e^{\mp2i\theta}\left(-\frac{\partial ^2}{\partial
r^2}\pm2ir^{-1}\frac{\partial ^2}{\partial r\partial
\theta}\mp2ir^{-2}\frac{\partial }{\partial\theta}
+r^{-1}\frac{\partial}{\partial r}+r^{-2}\frac{\partial ^2}{\partial
\theta^2}\right) \label{TETMPolar}
\end{eqnarray}
\end{widetext}

The non-zero coupling between the circular polarized components leads
to the mutual dependence of their winding numbers. The only
cylindrically symmetric solutions of Eqs.\ref{GPDimensionless} have
the following form:

\begin{eqnarray}
\left(\begin{array}{cc} \psi_+(r,\theta)\\ \psi_-(r,\theta)\end{array}\right)=e^{il\theta}\left(\begin{array}{cc} f_{+}(r)\\ e^{2i\theta}f_{-}(r)\end{array}\right)
\label{AnsatzLT}
\end{eqnarray}
which means that necessarily

\begin{equation}
l_+=l=l_--2
\label{Coupling}
\end{equation}

In terms of Ref. \cite{Rubo} this state correspond to a
winding number $k=-1$. Thus, one can conclude that the presence of the TE-TM splitting
does not allow the half-vortex as a stationary solution anymore.

The radial functions describing the vortex core can be found from the
following system of coupled equations, which can be obtained by
putting expressions \ref{TETMPolar}, \ref{AnsatzLT} into
Eq.\ref{GPDimensionless}.

\begin{widetext}
\begin{eqnarray}
\frac{1}{2}\left(\frac{d^2}{dr^2}+\frac{1}{r}\frac{d}{dr}\right)f_+-\left(A_1f_+^2+A_2 f_-^2-1+\frac{l^2}{2r^{2}}\right)f_++\chi\left(\frac{d^2}{dr^2}+\frac{2l+3}{r}\frac{d}{dr}+\frac{l(l+2)}{r^2}\right)f_-=0 \label{fplus}\\
\frac{1}{2}\left(\frac{d^2}{dr^2}+\frac{1}{r}\frac{d}{dr}\right)f_--\left(A_1f_-^2+A_2
f_+^2-1+\frac{(l+2)^2}{2r^{2}}\right)f_-+\chi\left(\frac{d^2}{dr^2}-\frac{2l+1}{r}\frac{d}{dr}+\frac{l(l+2)}{r^2}\right)f_-=0\label{fminus}
\end{eqnarray}
\end{widetext}
The above equations are quite complicated and only allow numerical
solution.

\section{Results and discussion}

In this section we present numerical results for radial
functions $f_\pm$ and the associated vortex polarization textures. To determine which configuration will have
the lowest energy, let us remind that without the TE-TM splitting, the
elastic energy of the vortex in a spinor condensate can be estimated
as \cite{Rubo}:

\begin{eqnarray}
E_{el}=\frac{\rho_s}{2}\int\left[(\nabla\theta_+)^2+(\nabla\theta_-)^2\right]d\textbf{r}\\
\nonumber\approx\pi\rho_s\left(l_+^2+l_-^2\right)\textrm{ln}\left(\frac{R}{a}\right)
\end{eqnarray}
where $\rho_s=\hbar^2n_\infty/m^*$ is the rigidity or stiffness of the condensate,
$a=\hbar/(m^*\mu)^{1/2}$ is the coherence length or the vortex core radius, $R$ is the size of the
system and $\theta_\pm$ are the phases of the circular polarized components.
From the above formula it follows that if $l_+=l_--2$, the minimal energy
corresponds to a vortex $(l_+,l_-)=(-1,+1)$. We thus start our analysis from
such a situation.

\begin{figure}[t]
\includegraphics[width=0.5\textwidth,clip]{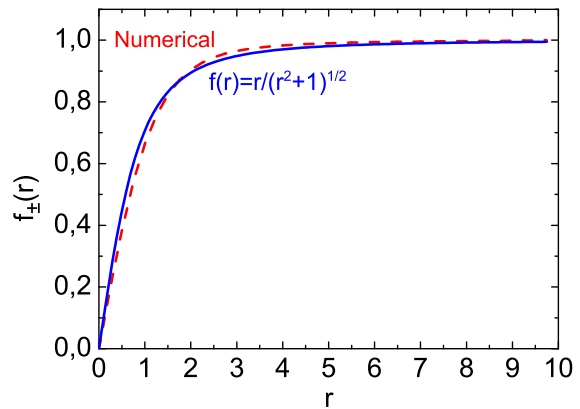}
\caption{(Color online) The exact numerical solution for radial
function of (-1,1) vortex (dashed red line) together with the
fitting function (solid blue line). The parameter values are
$\chi=-1/78$, $A_1=10/9$ and $A_2=-1/9$} \label{Fig1}
\end{figure}

In this case the radial functions corresponding to the opposite circular
polarizations found from numerical solution of Eqs.
\ref{fplus}, \ref{fminus} with $l=-1$ are identical and can be
satisfactory approximated by the following function plotted at Fig.\ref{Fig1} :

\begin{equation}
f_\pm(r)\approx\frac{r}{\sqrt{r^2+1}}
\label{Approx}
\end{equation}
One should make a remark at this point. Indeed, if $f_+=f_-$ is a solution of \ref{fplus} and \ref{fminus}, this is also the case for $f_+=-f_-$, but in such a situation the pseudospin will point in the opposite direction with respect to the first case. We will talk again about this in the next section.

\begin{figure}[t]
\centering
\includegraphics[width=0.5\textwidth,clip]{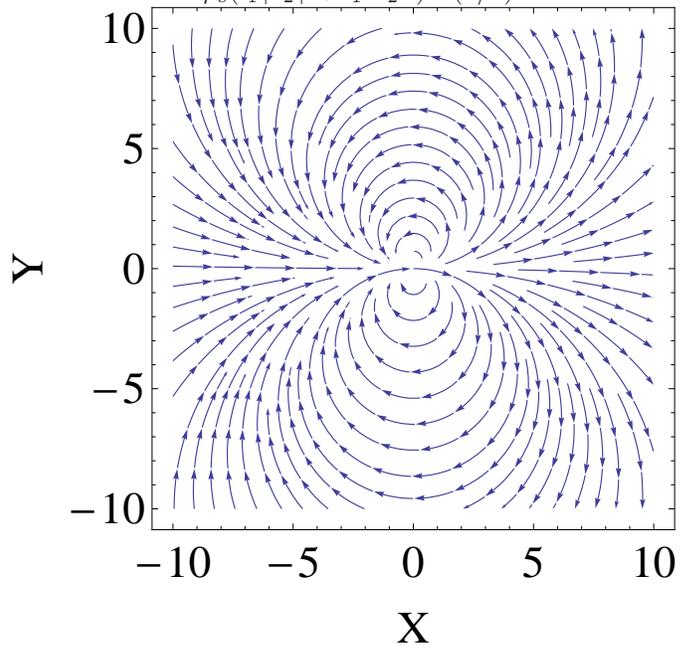}
\caption{(Color online) Pseudospin $(S_x,S_y)$ vector field for
$(-1,+1)$ configuration. This pattern is known to be the one of a
simple vortex with winding number 2. The pseudospin is aligned with
the TE-TM effective magnetic field} \label{Fig2}
\end{figure}

As $f_+=f_-$, the polarization of the system is always linear, which makes the
$z$-component of the pseudospin vanish : $S_z=0$.
The pseudospin lies in the plane and is at any point aligned with the TE-TM effective
field. The orientation of the TE-TM effective field depends on the wave vector orientation, which, in turn,
depends on the position of the particles with respect to the core of the vortex. The angular dependence both in
reciprocal and real space is given by the formula \ref{OmegaEff}, which shows that the orientation of the effective field
varies as 2 times the polar angle $\varphi$. The resulting polarization pattern
is the one of a simple (with only one winding number) vortex with winding number 2, as it's shown Fig.\ref{Fig2}.
If one investigates the $f_+=-f_-$ solution, the pseudospin will be totally symmetric and opposed to the TE-TM field,
which from one side would cost some energy, but which moreover would be a situation unstable against any perturbation.

Let us now try to understand qualitatively the way the TE-TM splitting would affect a half-vortex state, which
could be created, for instance, by some external means.
At $t=0$, $\sigma_-$ particles are almost homogeneously covering space and immobile. They are not affected by the TE-TM
splitting which is zero at $\overrightarrow{k}=0$. $\sigma_+$ particles are rotating. The pseudo-spin in the non-zero wave vector
states is fully aligned along z, perpendicular to the TE-TM field which is in the plane. The pseudo-spin therefore starts to rotate,
demonstrating that the half-vortex is not stationary. The speed of rotation is large close to the core where particles
rotate fast and where the TE-TM splitting is large, whereas the rotation is slower and slower going away from the center.
For each radius, the situation is reminiscent of the one happening in the optical Spin Hall effect \cite{OSHE}.
The density of $\sigma_+$, $\sigma_-$ particles is locally modified, which should provoke a drift of particles perpendicularly to the vortex motion,
and probably, a destruction of the vortex. The life time of such a transient state is therefore linked with the value of the TE-TM splitting
in the core region. We propose an estimation based on the value of the splitting $\beta_{\xi}$ seen by particles moving at the core radius $\xi$ characterized by a wave vector $\overrightarrow{k}_{\xi}$ :

\begin{equation}
\label{tau}
\tau  = \frac{\hbar }{{{\left|\beta _\xi\right|}\left( {{{\overrightarrow k }_\xi }} \right)}}
\end{equation}

This value can strongly depend on the type of structure, on the value of detuning etc.
The typical values which can be expected, however, lie between 10 and a few hundreds of picoseconds.
These times are comparable to the typical coherence times which have been measured for polariton condensates.
We conclude that the half-vortices could be experimentally observed both in resonant and non-resonant experiments.
They are however, intrinsically transient states with a life time probably limited by the TE-TM splitting
value. Thus they should be not considered in principle in a rigorous calculation of the BKT critical temperature.

The vortex $(-1,+1)$ can be considered as a bound state of two half-
vortices, $(-1,0)$ and $(0,+1)$. As it was shown in Ref.\onlinecite{Rubo}, without TE-TM splitting the interaction energy of
a pair of vortices $(l_{1+},l_{1-})$ and $(l_{2+},l_{2-})$ placed at
distance $d$ from each other reads:

\begin{eqnarray}
{E_{int}} \approx 2\pi {\rho _s}({k_1}{k_2} + {m_1}{m_2}){\rm{ln}}\left( {{a \mathord{\left/
 {\vphantom {a d}} \right.
 \kern-\nulldelimiterspace} d}} \right)\\\nonumber
=\pi {\rho _s}({l_{1 + }}{l_{2 + }} + {l_{1 - }}{l_{2 - }}){\rm{ln}}\left( {{a \mathord{\left/
 {\vphantom {a d}} \right.
 \kern-\nulldelimiterspace} d}} \right)
\end{eqnarray}

According to this formula, the half-vortices $(-1,0)$ and $(0,+1)$ do
not interact and the corresponding vortex pair is unbound. Let us reexamine the $(-1,+1)$ case, while adding the distance $d$ between $\sigma_+$ and $\sigma_-$ vortices along the $x$-axis. One has to write the associated wave function in cartesian coordinates with \ref{Approx}, $r=\sqrt{x^2+y^2}$ and $\theta_\pm=\arctan\left(y/x\right)+\pi H\left(-x\right)$ (H is the Heaviside function):

\begin{eqnarray}
\left(\begin{array}{cc} \psi_+(x+d,y)\\
\psi_-(x-d,y)\end{array}\right)=\left(\begin{array}{cc}
f_{+}(x+d,y)e^{il_+\theta_{+}(x+d,y)}\\ f_{-}(x-d,y)e^{il_-\theta_{-}(x-d,y)}\end{array}\right)
\label{VortPair}
\end{eqnarray}

\begin{figure}[t]
\centering
\includegraphics[width=0.5\textwidth,clip]{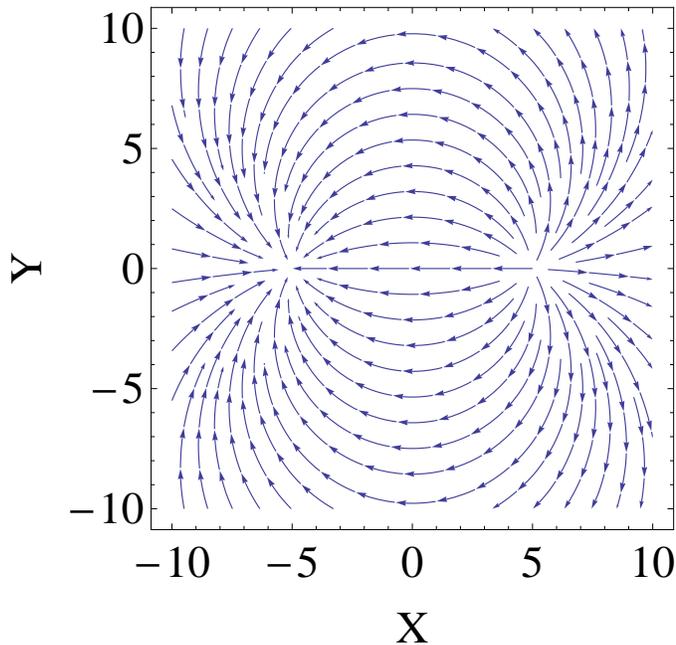}
\caption{(Color online) Pseudospin $(S_x,S_y)$ vector field for a
separation $d=5$ of the $\sigma_+$ and $\sigma_-$ vortices along the
$x$-axis.} \label{Fig3}
\end{figure}

The corresponding pseudospin configuration is shown Fig.\ref{Fig3}
for $d=5$ and the pattern compared
with the one of Fig.\ref{Fig2}
shows that as $d$ increases, the pseudospin becomes less and less aligned with the TE-TM field and the energy should consequently increase.
The normalized TE-TM energy part of the polariton condensate reads:
\begin{widetext}
\begin{eqnarray}
E_{TE-TM}=\rho_s\chi\int\left[\psi_{+}^*\left(\frac{\partial} {\partial y}+i\frac{\partial}{\partial x}\right)^2\psi_{-}+\psi_{-}^*\left(\frac{\partial} {\partial y}-i\frac{\partial}{\partial x}\right)^2\psi_{+}\right]d\textbf{r}
\label{ETETM}
\end{eqnarray}
\end{widetext}

\begin{figure}[t]
\centering
\includegraphics[width=0.5\textwidth,clip]{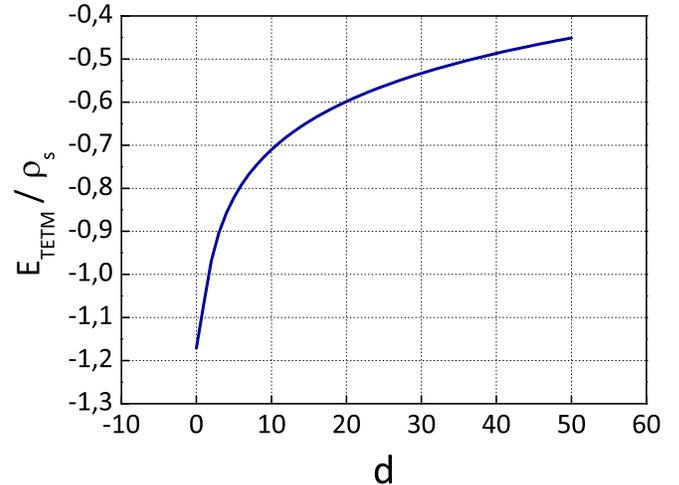}
\caption{(Color online) TE-TM normalized energy part as a function
of the separation $d$ for a $1000\times1000$ square system, the
lowest energy is reached for $d=0$}\label{Fig4}
\end{figure}

The numerical computation of $E_{TE-TM}$ as a function of $d$ is shown Fig.\ref{Fig4}.
One can see that the energy increases logarithmically with $d$ and that the lowest energy state is, as expected, the one with $d=0$, where the pseudo-spin is aligned with the effective field. Thus the TE-TM splitting makes $(-1,0)$ and $(0,+1)$ vortices interact and collapse on each other to form a $(-1,+1)$ vortex.
Let us remark that for the $f_+=-f_-$ solution, the opposite behavior will be observed and the energy will decrease when the vortices moved away from each other.
For $d=0$ and a system of size $\pi R^2$, the TE-TM and the kinetic energy read in polar coordinates:
\begin{eqnarray}
E_{kin} = \rho_s\frac{\pi }{2}\left[ {\frac{{{R^2}\left( {{R^2} - 2}
\right)}}{{{{\left( {1 + {R^2}} \right)}^2}}} + 2\ln \left( {1 +
{R^2}} \right)} \right]\label{Ek}\\
E_{TE-TM}= 2\chi E_{kin}
\end{eqnarray}

which with Eq.\ref{beta}, ${m^*}^{-1} = {2^{-1}}\left( {m_t^{-
1} + m_l^{ - 1}} \right)$ and the definitions of $\chi$ and $\rho_s$
gives ${E_c} + {E_{TE - TM}}=E_c^*$, where $E_c^*$ is the kinetic
energy associated with the new rigidity constant $\rho _s^* =
{n_\infty }{\hbar ^2}/{m_t}$. One concludes, as it could be expected, that the TE-TM effective
magnetic field switches the effective mass $m^*$ to the TE polarized
particles mass $m_t$ \cite{CommentMasses}.

\begin{figure}[t]
\centering
\includegraphics[width=0.5\textwidth,clip]{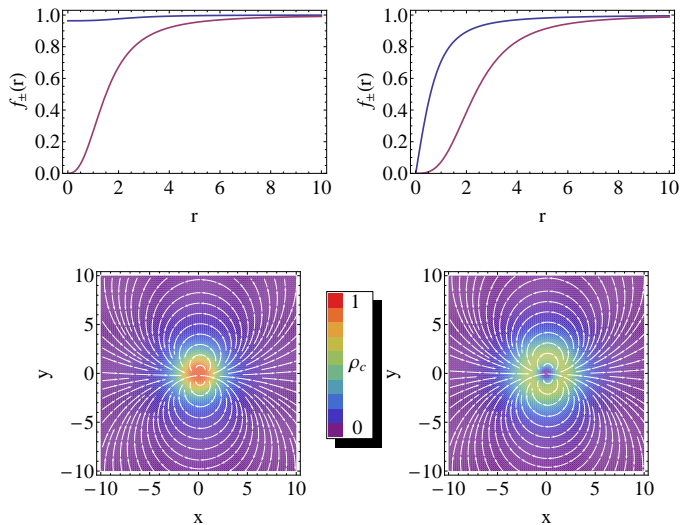}
\caption{(Color online) Left column is $(0,+2)$ and right column is
$(+1,+3)$ configuration. The top line shows radial functions with
a blue line for $f_+$ and a purple line for $f_-$. The bottom line
shows $\left(S_x(x,y),S_y(x,y)\right)$ vector fields (white arrows)
over $\rho_c=2S_z(x,y)$ (the degree of circular polarization) background.}\label{Fig5}
\end{figure}

Finally we will say a word about $(0,+2)$ and $(+1,+3)$ configurations that, if they are not energetically favorable,
exhibit interesting pseudospin (polarization) patterns. One can note by the way, that these two states are totally symmetric respectively to $(-2,0)$ and $(-3,-1)$. In these cases radial functions are no more identical for the two components, which will introduce a nonzero circular polarization close to the vortex core.
Numerically calculated radial functions are plotted in the upper part of Fig.\ref{Fig5}
and $S_z(x,y)$ functions as the background of $(S_x,S_y)$ vector fields at the lower part. One can remark that the latter vector fields are exactly the same as the one of Fig.\ref{Fig2}. Indeed, this configuration is fixed by the condition \ref{Coupling}.
The $(0,+2)$ state is peculiar, so far as there is no vortex for $\sigma_+$. Nevertheless, the corresponding radial function is not constant as expected. Indeed, the interaction between circular components implies a small depletion around the center of the system, observed as a minimum at $r=0$. The polarization becomes more and more circular while approaching $r=0$, but is never fully circular.
In the $(+1,+3)$ configuration, one has a vortex for each component and the pseudospin $S_z$ component exhibits a maximum before reaching $r=0$ which corresponds to a ring around the vortex core that figures out the maximum of circular polarization degree at about $r=0.6$.

\section{conclusions}

In conclusion, we analyzed the impact of the TE-TM splitting on vortices
in spinor polariton condensates. We have shown that this splitting induces a qualitative change
of the nature of the stationary vortex state supported by a polariton condensate.
The half-vortices are no more stationary solutions of the spinor Gross-Pitaevskii
equations and should not affect the critical temperature of the BKT phase transition.
Their life time is of the order of 10 to a few hundreds of ps, limited by the TE-TM splitting value.
However, they can, in principle, be observed experimentally.
The stable vortex having the smallest energy is the state $(-1,+1)$ (in the circular basis), whose
polarization pattern follows the one implied by the peculiar TE-TM symmetry.
Polarization textures of other vortex states ($(0,+2)$ and $(+1,+3)$) have also been analyzed.


\begin{thebibliography}{99}
\bibitem{CommentMasses} We supposed that $m_t>m_l$ and thus $\chi<0$. In the case of the vortex in which the pseudospin is oriented antiparallel to effective magnetic field, one should replace $m_t$ by $m_l$.
\end{thebibliography}
\end{document}